\documentclass[%
aps,prl,twocolumn,showpacs,superscriptaddress
]{revtex4-1}

\usepackage{graphicx}
\usepackage{dcolumn}
\usepackage{bm}
\usepackage{amssymb,amsfonts,amsmath}
\usepackage{amsmath}
\usepackage{pifont}
\usepackage{amsfonts}
\usepackage{color}

\begin{document}

\title{Why Microtubules run in Circles - Mechanical Hysteresis of the Tubulin Lattice}

\author{Falko Ziebert}
\affiliation{Albert-Ludwigs-Universit\"at, 79104 Freiburg, Germany}
\affiliation{Institut Charles Sadron UPR22-CNRS, 67034 Strasbourg, France}
\author{Herv\'{e} Mohrbach}
\affiliation{Groupe BioPhysStat, LCP-A2MC, Universit\'e de Lorraine, 57078 Metz, France}
\affiliation{Institut Charles Sadron UPR22-CNRS, 67034 Strasbourg, France}
\author{Igor M. Kuli\'{c}}
\affiliation{Institut Charles Sadron UPR22-CNRS, 67034 Strasbourg, France}


\hyphenation{GMPCPP}

\begin{abstract}
The fate of every eukaryotic cell subtly relies on the exceptional
mechanical properties of microtubules. Despite significant
efforts, understanding their unusual mechanics remains elusive. 
One persistent, unresolved mystery is the formation 
of long-lived arcs and rings, e.g.~in kinesin-driven gliding assays.
To elucidate their physical origin 
we develop a model of the inner workings of the
microtubule's lattice,
based on recent experimental evidence
for a conformational switch of the tubulin dimer.
We show that the microtubule lattice itself coexists in
discrete polymorphic states. 
Curved states can be induced via a mechanical hysteresis involving 
torques and forces typical of few molecular motors acting in unison.
This lattice switch renders microtubules not only
virtually unbreakable under typical cellular forces, but
moreover provides them with a tunable response
integrating mechanical and chemical stimuli.
\end{abstract}

\pacs{87.16.Ka, 82.35.Pq, 87.15.-v}
\maketitle


Microtubules (MTs) are the stiffest cytoskeletal component and
play many versatile and indispensable roles in living cells.
As 'cellular bones', they define to a large part cell mechanics,
and are crucial for cellular transport 
and cell division \cite{Howardbook,Amosbook,MTStirringRod,spindle}.
Beyond their biological importance, MTs have been used as
molecular sensors for intracellular forces,
as biotemplates for nanopatterning,
and as building blocks for 
hybrid materials
and active systems like artificial cilia
and self-propelled droplets \cite{Gijse_sens,Meyrhoef,Diez_templ,DiezSM,Dogic_cil,Dogic_motion}.
The MT's structure is well known \cite{Nogalesrev}:
the elementary building blocks, tubulin dimers,
polymerize head to tail into linear protofilaments,
that associate side by side to form the
hollow tube structure known as the MT.

\begin{figure}[b]
\includegraphics[
width=0.98\linewidth]{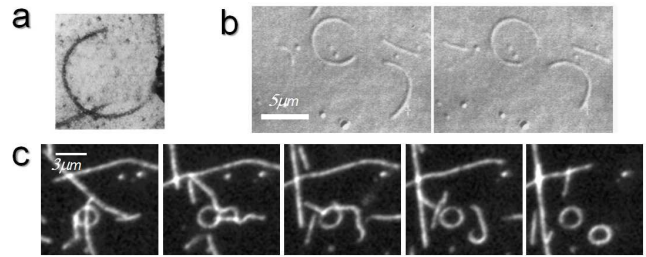}
\caption{(a)-(c) Mysterious 
ring formation
of microtubules gliding on a kinesin motor carpet. 
From \cite{Amos},\cite{Boehm} and
\cite{Ross}, respectively.
}%
\label{fig1}%
\end{figure}

Despite of the MTs'
importance,
widespread use, the knowledge of its structure, and
numerous experiments probing their elastic properties
\cite{MTBending,MTBending2,MTBending3,Pampaloni,Taute},
understanding their basic mechanics still poses challenging
problems.
A remarkable one 
is found in 
MT gliding assays
\cite{GlidingAssay1,Amos,Vale,Bourdieu}, see Fig.~\ref{fig1}:
already two decades ago, Amos \& Amos \cite{Amos} observed that
MTs driven by kinesin motors on a glass surface
can form 
arcs which continue gliding for significant time intervals before
suddenly straightening out. They remarked with
quite some foresight that these circular MT states could 
be explained by the
existence of alternative tubulin dimer conformations
\cite{AmosReview}.
The observation remained without wider public notice despite
the frequent reoccurrence of MT arcs
in the gliding dynamics of single filaments
\cite{Boehm,Surrey,Ross}, bundles
\cite{Hess1,Hess2,Kawamura} and in collective
(high density)
gliding 
\cite{chate}.
Force-induced circular arcs on the same scale,
but rather dissimilar to classical 
buckling, have been found in numerous other
situations \cite{Vale,Koch}, also {\it in vivo}
\cite{Borisy,Odde,Brangwynne}.
While the lifetime of MT rings varies,
their characteristic size of about one micron
is preserved in single filament experiments
\cite{Amos,Boehm,Surrey,Ross}, indicating a 
robust mechanism at work.
The importance of internal degrees of freedom of the
MT lattice (e.g.\,inter-protofilament shear
\cite{SoftShearModel1,SoftShearModel2} and tubulin conformational
switching \cite{Mohrbach2010})
has been recently stressed in explaining another MT `anomaly',
the surprising length dependent stiffness in clamped MT experiments
\cite{Pampaloni,Taute}.
However, the formation of long
lived, highly curved states or rings could not be rationalized so
far. We here present a model of MT mechanics
that integrates the current experimental knowledge. 
It shows that under
external forces, MTs can be converted into metastable circular states,
explaining the recurrent
observations of gliding rings.

\begin{figure}[t]
\includegraphics[width=0.98\linewidth]{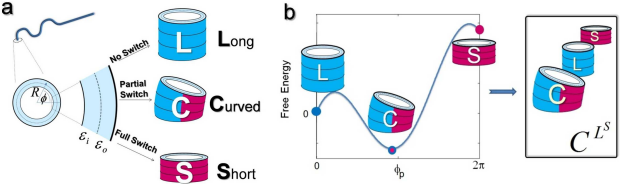}
\caption{ 
(a) The bistable tube model with
internal prestrains leading to three different lattice states:
short and straight ("S"), long and straight ("L") and curved
("C"). (b) A typical polymorphic energy landscape,
cf.\,Eq.\,(\ref{EPhip}), with three metastable states
and the associated polymorphic signature (with C the ground
state, L and S "excited" states).
}%
\label{fig2}%
\end{figure}

{\it Switchable Tube Model.}
We will develop a simple yet generic elastic model of MTs
that principally revises their mechanical response.
We make use of the following
experimental facts about tubulin 
and the MT lattice \cite{NOTEtax}. 
Fact 1: The 
protofilament can
coexist in at least two conformational states
\cite{Multistable_Tub_EM}, one of them 
straight, the other one highly curved with radius of curvature
$20$-$30$ nm 
\cite{ref20nm,Multistable_Tub_EM,NOTEWeakluCurvedState}. Fact 2:
The 
MT lattice displays two
elongational states with one of them being
about $2\%$
shorter \cite{ArnalWade}. 
The MT is modeled, cf.\,Fig.~\ref{fig2}a,
as a bi-layered tubular structure with an inner and an
outer layer of material both together representing the
coarse-grained lattice of tubulin dimers.
We parameterize the MT cross-section by the
azimuthal angle $\phi$ and the radial distance $R$
to the centerline. 
For a MT containing $N$ protofilaments (typically $13$),
a circumferential block of $p$
consecutive dimers is hence given by
$\phi\in\left[0,\frac{2\pi p}{N}\right]$,
$R\in\left[R_{i},R_{o}\right]$, with $R_{i,o}$ the inner and
outer MT radii known from crystallography \cite{Howardbook}.
If the tube were to be intrinsically
relaxed and preferentially straight, once it is bent or stretched
it will endure inner strains, in the simplest case
via a combination of pure stretching and
geometric curvature contributions, $\varepsilon\left(
R,\phi\right) =-\vec{\kappa}\cdot\mathbf{R}+\bar{\varepsilon}$.
Here $\vec{\kappa}$ 
is the vectorial curvature and $\bar{\varepsilon}$ the mean stretching
strain of the cross-section.
The conformational `polymorphic' state of a tubulin is described by a
variable $\sigma$, assuming one of two states:
straight, with $\sigma=0$ and vanishing preferred
strain $\varepsilon_{pref}=0$, and curved
with $\sigma=1$ generating finite prestrain.
In general the prestrain has different values
$\varepsilon_{pref} =\varepsilon_{i}$ or $\varepsilon_{o}$
in the two layers, see Fig.~\ref{fig2}a.

The energy density of the MT's cross-section
consists of two parts: the elastic energy (with $Y$ Young's modulus) 
$e_{el}
=\frac{Y}{2}\int_{R_{i}}^{R_{o}}dr\int_{0}^{2\pi}r\,d\phi\left[
\varepsilon\left(r,\phi\right)-\varepsilon_{pref}\left(r,\phi\right)
\right]^{2}.
$
Second, each dimer
can reduce its free energy by $|\Delta G|$ by switching to the curved
state. 
The respective energy of the cross-section is
$e_{switch}=\frac{\Delta G}{b}\sum\nolimits_{n=1}^{N}\sigma_{n}$ with
$b\approx8\,\mathrm{nm}$ the tubulin size.
A negative free energy, $\Delta G<0$, will favor the curved state.
Within the lattice, however, dimer
switching 
competes with the elastic energy. 
A tube section with a block of dimers switched to the curved state 
will endure prestress that can be (partially)
relieved by global tube deformations:
the MT will 
show axial shortening/lengthening
and, remarkably for the observer, 
curving. External forces, $F$, and torques, $M$,
give rise to additional couplings.
The total energy $e_{tot}=e_{el}+e_{switch}$ can be calculated to be \cite{suppref}
\begin{equation}\label{en_before_min}
\tilde{e}_{tot}
=\frac{{\kappa}^{2}}{2}+\frac{a_{0}\bar{\varepsilon}^{2}}{2}
-{\kappa}\sin\frac{\phi_{p}}{2}+a_{1}\phi_{p}\bar{\varepsilon}
+m{\kappa}+\lambda\bar{\varepsilon}+\gamma\phi_{p}\,,
\end{equation}
where we scaled curvature by its characteristic value
$\kappa_1$, energy by $B\kappa_{1}^{2}$,
torque like $m=M/B\kappa_{1}$
and tension like $\lambda=F/B\kappa_{1}^{2}$.
The newly introduced variable
$\phi_{p}=2\pi p/N$
is the angular size of the switched block, to which we refer to as
the `polymorphic variable' in the following.
It is this inner variable of the lattice, that is often concealed from
observation
but can give
rise to surprising effects.
The characteristic curvature $\kappa_1$ is completely determined
by the known tube dimensions and the prestrains.
Using the experimental facts 1 \& 2,
we estimate the preferred strains to be
$\varepsilon_i=0.7\cdot10^{-2}$ and
$\varepsilon_o=-3.3\cdot10^{-2}$, see \cite{suppref}
for details. $\kappa_1$ will set
the scale for the gliding arcs and rings.
As an important cross-check, using the estimated strain values
we get
$\kappa_1\simeq 1.1 \mu{\rm m}^{-1}$,
consistent with the experimental observations 
\cite{Amos,Boehm,Surrey,Ross}.
The first two terms in Eq.\,\ref{en_before_min} are of purely
elastic origin, the next two are cross-coupling terms between
elastic and polymorphic variables, and the last three
are related to the action of generalized forces. 
$\gamma=a_{2}+\frac{N\Delta G}{2\pi b B\kappa_{1}^{2}}$ is the
effective energy density of switching; it consists of an elastic
penalty due to the lattice constraint, $a_{2}$,
and of the free energy difference per length, $\Delta G/b$,
due to switching. $a_{0},a_{1}$ and $a_{2}$ are completely determined dimensionless
constants \cite{suppref}.

To carve out the features of this energy,
we first consider 
imposed 
external force and torque.
A minimization with respect to
strain $\bar{\varepsilon}$ and curvature $\kappa$
yields
\begin{equation}
{e}_{tot}\left(  \phi_{p}\right)  =-\frac{c_{p}}{2}\phi_{p}^{2}
+f\phi_{p}-\allowbreak m\sin\frac{\phi_{p}}%
{2}-\frac{1}{2}\sin^{2}\frac{\phi_{p}}{2}\,, \label{EPhip}%
\end{equation}
a function of torque $m$ and
effective tension
$f=\gamma-c_{f}\lambda$.
For the given prestrains $\varepsilon_{i,o}$
we estimate the dimensionless constants $c_{p}\simeq0.1$, $c_{f}\simeq2.5\cdot10^{-3}$,
and $\gamma$ between 0 and 1 for $|\Delta G|$ of the order of several kT, 
see \cite{suppref} for details. The generalized force is
a function of lattice geometry and tension, and
most importantly of the switching energy, i.e.\,$f=f(\Delta G)$.

Eq.\,(\ref{EPhip}) is the central result: it describes the energy of a
MT cross-section with partially switched tubulins and
contains all the information
needed to characterize the mechanical behavior of MTs
under external loads. As a function of the acting force and torque,
the `state diagram' of a MT cross-section displays a
number of distinct regions: the energy 
landscape can exhibit one, two or three
local minima, cf.\,Fig.\,\ref{fig2}b. The states corresponding to
these minima are a long straight state (denoted with "L") 
with $\phi_p=0$, a short straight state ("S") 
with $\phi_p=2\pi$ and a curved state ("C") having an intermediate
value $0<\phi_p<2\pi$. For further discussion 
we introduce an intuitive notation (\emph{polymorphic
signature}) capturing both the energy shape 
and the actual state, 
taking one of these forms: a single
mechanically stable state $X$, two  $X^{Y}$ or three
stable states $X^{Y^{Z}}$  (with $X,Y,Z$ either one of $S,L,C$).
Increasing energy values are indicated by ascending indices
and the actual state by underlining the
corresponding index. For instance, in
$\underline{X}^{Y^{Z}}$ the ground state $X$ is populated, while in
$X^{\underline{Y}^{Z}}$ metastable state $Y$.

\begin{figure}[t]
\includegraphics[width=0.98\linewidth]{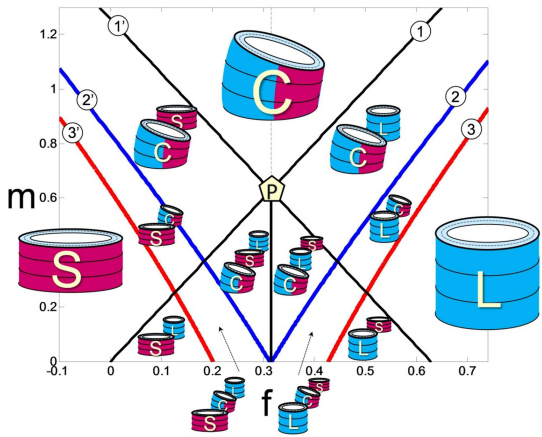}
\caption{The polymorphic state diagram for a MT cross-section
as a function of  effective force
$f$ and  torque $m$, cf.\,Eq.\,(\ref{EPhip}).
The diagram depends on the parameter $c_p$, estimated 
from the experimental measurements of
protofilament curvature and MT lattice shortening (cf.\,main text).
}%
\label{fig3}%
\end{figure}

The complete variety of polymorphic signatures physically allowed
by ${e}_{tot}$
for given torque $m$ and effective force $f$ 
is summarized in the `polymorphic state diagram' in
Fig.\,\ref{fig3}.
Note that the simple model -- considering two dimer states -- 
leads to an extremely rich lattice behavior.
There is an intrinsic symmetry with respect to the vertical line given by $f=\pi c_p$
and passing through the point P: a mirror operation with respect
to this line transforms the two straight states $S$ and
$L$ into each other, leaving the curved state invariant.
We will hence restrict the discussion to MTs initially in the $L$
state. The most interesting, `polymorphically curved region',
i.e.~the range where stable or metastable curved states occur, is
comprised between the two red curves, \ding{174} and
\ding{174}$'$. Outside of these curves, the behavior is {\it
indistinguishable} from the one of a simple elastic beam or
worm-like chain (WLC). Between the two blue curves, \ding{173} and
\ding{173}$'$, the curved state $C$ has lowest energy. Moreover,
in the region above the point $P$ and between the two black lines
\ding{172} (given by $m=2f$) and \ding{172}$'$, the curved state
$C$ is even the {\it only} existing state. Finally, MTs are most
senstive to 
external loads close to
point $P$, where the curved state is the ground state in all
surrounding regions.
One expects `polymorphic behavior' --
i.e.\,(parts of) MTs curved as in Fig.\,\ref{fig1} --
to become visible when the respective cross-sections are mechanically
converted to the C state. From the state diagram the necessary torque for this conversion,
depending also on the force, can be estimated to be of the
order of $10\,{\rm pN}\mu{\rm m}$ (corresponding to $m=1$ in
reduced units). As a single kinesin motor exerts forces of
$2$-$5\,{\rm pN}$ \cite{Howardbook} on MTs with typical
radius of curvature of $1 \mu m$, this provides us
with a first clue for understanding the MT's propensity
to form rings in gliding assays, even when motor coverage
on the substrate is moderate.

\begin{figure*}[t]
\includegraphics[width=7in]{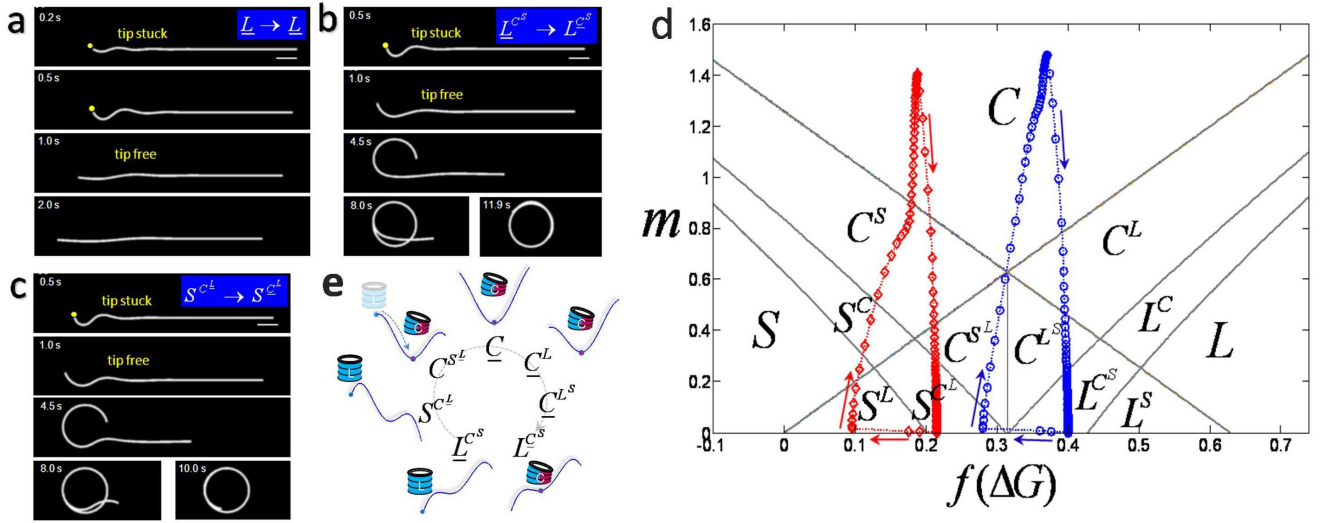}
\caption{ Dynamic shape evolutions of gliding MTs that
become temporarily stuck and buckle. All parameters are identical
in a)-c), except for $\Delta G$ resulting in
different 
effective forces $f$, cf.\,Fig.\,\ref{fig3}
and panel d.
a) A MT far from
the polymorphic region ($\Delta G= +1.7$ kT
corresponding to $\gamma=1$) behaves like a gliding WLC upon buckling and release.
b) A MT for which the buckling
induces a switch to the curved state ($\Delta G= -5.2$ kT,
$\gamma=0.4$). The finally formed ring is metastable, i.e.\,$L$ has
lower energy than $C$.
c) A MT displaying multiple
buckles that persist after unsticking and roll-up ($\Delta G=
-7.3$ kT, $\gamma=0.215$).
d) Trajectories in the polymorphic state diagram
corresponding to the dynamics shown in b (blue curve) and c (red
curve), resulting in a hysteretic loop upon torque creation and release,
cf.\,the sketch in e).
Scale bars in a)-c) are $1\mu {\rm m}$, gliding velocity
$v=1\,\mu{\rm m/s}$. All parameter values can be found in \cite{suppref}.
}%
\label{fig4}%
\end{figure*}

{\it Gliding-induced Ring Formation.} To investigate whether the
found polymorphic signatures, emerging from the switchable
internal structure of the MT, indeed explain ring formation, we
generalized previous gliding assay modeling
\cite{Bourdieu,Nedelec} and simulated buckling events of gliding
MTs. Due to the confinement by the motors, the MT can be described
by a 2D space curve $\mathbf{r}(s)$. We assume that the MT's shape
and the inner polymorphic variable $\phi_p(s)$ both follow
relaxational dynamics
\begin{equation}
\dot{\mathbf{r}}
=\left(  \frac{\mathbf{n}\mathbf{n}}{\xi_{\perp}}
+\frac{\mathbf{t}\mathbf{t}}{\xi_{\parallel}}\right)\hspace{-.5mm}
\cdot
\hspace{-.5mm}
\left[  -\frac{\delta E}{\delta\mathbf{r}}+f_{m}\mathbf{t}\right],\,\,
\dot{\phi}_{p}=-\frac{1}{\zeta}\frac{\delta
E}{\delta\phi_{p}}\,,\label{eqmot}%
\end{equation}
given by the polymorphic energy
\begin{equation}\label{energynumerics}
E=\hspace{-1mm}\int_{0}^{L}\hspace{-2mm}ds\left[ B\kappa_1^2\tilde{e}_{tot}+\frac{1}{2}\left(  \mathbf{t}^{2}\hspace{-1mm}-1\right)  \lambda
+\frac{B_{p}}{2}(\partial_s\phi_{p})^{2}\right],
\end{equation}
augmented by a length
constraint (introducing the Lagrangian multiplier $\lambda(s)$)
and a term penalizing variations in the polymorphic
variable along the arc length. The latter introduces a polymorphic
stiffness parameter, $B_{p}$, modeling a certain cooperativity in
lattice switching 
\cite{Mohrbach2010,Mohrbach2012}.
In the dynamic equations, $\xi_{\perp}$ and $\xi_{\parallel}$ are anisotropic friction coefficients
related to motor friction, and
$\zeta$ is associated to dissipation in the 
switching.
$f_{m}$ is the local force density exerted by the kinesin motors
attached to the substrate, transporting the filament along the
local tangent $\mathbf{t}$.
More details 
can be found in \cite{suppref}.

What happens if an initially straight MT
gets stuck and buckles due to the action of
motors?
The answer depends on torque and force, varying in
the course of buckling as follows (cf.\,blue curve
in Fig.\,\ref{fig4}d):
initially, the MT
is gliding freely, $f=\gamma$. Upon getting stuck,
compressive tension 
is rapidly created
leading quasi-instantaneously to
$f=\gamma-c_{f}\lambda$, 
followed by a slower build-up of torque due to
the emerging curvature. As long as the MT tip remains stuck
the torque increases while tension partially
relaxes via buckling, see the part of the blue curve in
Fig.\,\ref{fig4}d with the arrow pointing upwards.
Once the MT tip is released, the torque relaxes and one finally
ends up at the starting point at $f=\gamma$ and $m\simeq0$.
This 'loop' in the 
state diagram broadens and extends to higher torque values with
increasing motor force $f_m$. Depending on the effective tension and torque
induced during buckling, several distinct scenarios are possible:

For negligible dimer switching
energies $\Delta G \simeq 0$ kT ($\gamma \simeq 1$) or 
positive ones, a MT initially in the $L$ state buckles
as a classical semiflexible filament (WLC) as shown in
Fig.\,\ref{fig4}a:\,buckles
with force-dependent curvature emerge close to the tip and
straighten again after unsticking.
The same behavior is found for strongly
negative values of $\Delta G\lesssim-15$ kT ($\gamma \simeq-0.4$), and
a MT starting out from state $S$.
In both cases, the induced torques are insufficient for
crossing curve \ding{172} or \ding{172}$'$, consequently the
MT remains in the stable $L$ (or $S$) state.

For intermediate values of
dimer switching energies $-6.3$ kT$\lesssim\Delta G\lesssim-5.1$
kT ($0.3\lesssim\gamma\lesssim0.4$) the critical line \ding{172}
is crossed for high enough torques. Consequently the $C$ state
becomes populated at the filament tip, resulting in arc-like buckles
with curvature $\kappa\simeq\kappa_1$, see Fig.\,\ref{fig4}b.
Upon unsticking and concomitant torque
decrease, state $L$ acquires lowest energy. Remarkably, despite of
$L$ being the ground state, the leading buckle locks in the $C$
state, i.e.\,in the metastable configuration
$L^{\underline{C}^S}$. This scenario of a closed loop in the polymorphic state diagram
triggered by external forces and
ending up in a metastable curved state, 
could be called {\it `mechanical hysteresis'}.
The related path 
is shown as the blue curve in Fig.\,\ref{fig4}d and the concomitant changes
in the energy landscape and the polymorphic signatures passed
in Fig.\,\ref{fig4}e. Although state $C$ is only
metastable, the whole MT eventually rolls up and converts
to $C$ by simply following the motion of the tip in the course of
the gliding dynamics, as shown in Fig.~\ref{fig4}b.
For  dimer switching energies $-7.5$ kT $\lesssim\Delta G\lesssim-6.3$ kT
($0.2\lesssim\gamma\lesssim0.3$) the behavior is similar as just discussed,
when the initial state $L$ is replaced by $S$.
The MT 
converts into the $C$ state upon
buckling. The polymorphic signature will evolve from 
$\underline{S}^{{C}^L}$ into $S^{\underline{C}^L}$.

Finally, Fig.\,\ref{fig4}c shows yet another scenario, 
that takes place when 
$S$ is the ground state, yet the
MT starts from the $L$ state (red curve in Fig.\,\ref{fig4}d).
Due to the polymorphic landscape, 
state $C$ is populated almost
immediately during buckling, 
buckles remain stable after unsticking
and the final state is again a ring, but
in state $S^{\underline{C}^L}$.

The most sensitive parameter controlling
polymorphic behavior is the free energy difference $\Delta G$:
changes in the range from 0 to -10kT fundamentally alter the
response to external forces.
This suggests deep
impact of various binding agents,
like the GTP-analogue GMPCPP \cite{Vale},
MKAC \cite{Multistable_Tub_EM} or tau
\cite{TauBindingToCurvedMT}, either triggering or inhibiting
polymorphic behavior.
Our model yields estimated bounds for $\Delta G$:
if a MT forms metastable rings 
our analysis implies that 
$-7.5$ kT $\lesssim\Delta G\lesssim-5$ kT.

{\it Conclusions.} 
We have developed a generic model that principally revises
the mechanics of microtubules and explains their metastable curved states.
The conceptually simple model
comprises  the classical semiflexible filament behavior
at low loads. However, when the applied forces and torques exceed
a threshold (tens of pN), MTs 
convert to 
metastable curved conformations via a mechanical hysteresis loop.
The mechanically induced 
polymorphic switching shares similarities with bacterial
flagella \cite{Berg,Powers,Netz}. The metastability 
-- MTs "remember" force-induced curved shapes --
puts them close to man-made shape memory materials \cite{ShapeMemory}.
It is intriguing that already two
states of the subunit give rise to a plethora of polymorphic
signatures that can be triggered externally.
The strong sensitivity to the 
transition energy shows that the design is highly versatile and
could inspire novel smart materials. The major
challenge will now be to unravel how the microtubule
polymorphic switch is utilized in Nature.

\begin{acknowledgments}
{\it Acknowledgements.} We thank A. Johner, K.J. B\"ohm and D.
Chr\'etien for stimulating discussions and A. Maloney, L.
Herskowitz, and S. Koch for publishing and sharing their gliding
assay image series open data. F.~Z.~thanks the DFG for partial
support via IRTG 1642 Soft Matter Science.
\end{acknowledgments}


\begin{thebibliography}{37}
\bibitem {NOTEtax}
These facts were
established for taxol stabilized protofilaments/MTs, a
protocol also used in the majority of the gliding
experiments.
\bibitem {NOTEWeakluCurvedState} Note that in \cite{Multistable_Tub_EM} a third, less curved state
with radius of curvature of $250\,$nm was also detected, which we neglect here 
as the MT shape will be dominated by the highly curved state.
\end{thebibliography}
\end{document}